\documentclass[conference, letterpaper]{IEEEtran}
\usepackage{graphicx}
\usepackage{placeins}
\usepackage{wrapfig}
\usepackage{array}
 
\usepackage[cmex10]{amsmath}
\usepackage{color}
\usepackage{fancyhdr}
\usepackage[caption=false,font=footnotesize]{subfig}

\fancypagestyle{plain}{
        \fancyhead{}
        \fancyhead[C]{first page center header}
        \fancyfoot{}
        \fancyfoot[C]{first page center footer}
}

\begin{document}
\raggedbottom

\title{Survey of Automated Vulnerability Detection and Exploit Generation Techniques in Cyber Reasoning Systems}

\author{\IEEEauthorblockN{Teresa Nicole Brooks}
\IEEEauthorblockA{Seidenberg School of Computer Science and Information Systems\\ 
Pace University, New York NY\\
Email: tb93141n@pace.edu}}

\maketitle

\begin{abstract}
Software is everywhere, from mission critical systems such as industrial power stations, pacemakers and even household appliances. This growing dependence on technology and the increasing complexity of software has serious security implications as it means we are potentially surrounded by software that contains exploitable vulnerabilities. These challenges have made binary analysis an important area of research in computer science and has emphasized the need for building automated analysis systems that can operate at scale, speed and efficiency; all while performing with the skill of a human expert. Though great progress has been made in this area of research, there remains limitations and open challenges to be addressed. Recognizing this need, DARPA sponsored the Cyber Grand Challenge (CGC), a competition to showcase the current state of the art in systems that perform; automated vulnerability detection, exploit generation and software patching.  This paper is a survey of the vulnerability detection and exploit generation techniques, underlying technologies and related works of two of the winning systems Mayhem and Mechanical Phish.

\end{abstract}


\section{Introduction}
Technology touches every aspect of our lives, from the mundane to mission critical systems that facilitate our very way of life. These facts present clear economic, safety and security concerns. These concerns are driving the need for automated, scalable and reliable means of discovering, verifying and patching exploitable defects. In an effort to drive research in this area, DARPA sponsored the Cyber Grand Challenge (CGC), a competition to showcase the current state of the art in Cyber Reasoning Systems. These systems combine various tools, techniques and expert knowledge to create fully autonomous systems that perform automated vulnerability detection, exploit generation and software patching in binary software without human intervention. In this competition competing systems play an ``attack-defend'' style of Capture The Flag (CTF). CTF is ``a head-to-head, networked competition'' where participants are to detect, patch and exploit software defects \cite{DARPA}.

\subsection{Impact of DARPA's Cyber Grand Challenge}
In other areas of computer science research that involve the development of intelligent systems, such as machine learning and artificial intelligence there is a wealth of common datasets and corpora with corresponding benchmarks by which researchers can evaluate the efficacy of their approaches in a platform and technology agnostic way. An example of such a dataset is the ``MNIST database of handwritten digits.'' This dataset has a rich history of benchmarks and provides a standard dataset for training neural networks and other machine learning algorithms \cite{LeCun}\cite{LeCuna}. However, in the field of security research, specifically the areas of binary analysis such datasets and benchmarks do not exist. This often means that techniques are evaluated on different datasets (software) and different platforms, thus making it difficult to compare the effectiveness of different techniques \cite{Shoshitaishvili}.

DARPA's Cyber Grand Challenge addresses this need for a common platform and datasets by which to evaluate Cyber Reasoning Systems. CGC organizers designed binaries called challenges that differ in complexity, file size and functionality. These binaries are designed to present the same challenges of real-world software to the systems analyzing them. This collection of binaries coupled with a Linux distribution designed for the competition called DECREE OS, offers a standard platform and dataset for all competitors to evaluate their systems.  The qualifying round results, binaries, environment, needed libraries and documentation have all been made freely available online. This provides benchmarks and a common platform for researchers to test the effectiveness of new analysis techniques and systems \cite{DARPAa}.

Systems are judged based on security, availability and evaluation. Patched binaries (challenge replacement binaries) functionality is tested by running tests created by the CGC organizers. These tests are in the form of proof of vulnerability (POV). If no POVs are blocked their security score is 0. Patched binaries are also rated on their overhead on system resources such as memory, CPU usage and the file size. Table I shows a summary of the scoring criteria for competing systems. Note, systems that submit a working POV along with their patched binary have their security score doubled. Further note, the scoring algorithm suggests a stronger emphasis on binary patching versus the number of exploitable defects found by the competing systems \cite{Summary2014}.

\subsection{Limitation of The Study}
Although this work intends to survey the automated vulnerability detection and exploit generation techniques of current state of the art Cyber Reasoning Systems, there are gaps in this research. These gaps exist because many of the competing systems in DARPA's Cyber Grand Challenge were purpose built or at the very least were augmented versions of their original design and implementations in order to meet the criteria for the competition; thus making it very difficult to find literature to write an exhaustive survey. To address these limitations, future work includes collaborating with researchers who designed and implemented these state of the art systems in order to produce a more comprehensive survey.

\begin{table}
\centering
\begin{tabular}{ |p{2cm}|p{5cm}| } 
\hline
\multicolumn{1}{|c|}{\textbf{Criteria}} &  
\multicolumn{1}{c|}{\textbf{Rule}}\\
\hline
Security & Each competitor can defend the code on its server, keeping flags safe. It can patch each challenge binary using generic defenses or a custom patch for each vulnerability it finds.\\
\hline
Availability & Every program on a server should function normally after being patched as it would be easy to defend software if you could just disable all its functionality. The reference checks that defended software is responding correctly and hasn't been disabled or slowed.\\
\hline
Evaluation & Every player can program a vulnerability scanner. Searching for vulnerabilities in opponents software and proving weaknesses to the referee. A successful proof counts as a captured the flag.\\ 
\hline
\end{tabular}
 \bigskip
\caption{Summary of CGC scoring rules \cite{Summary2014}}
\end{table}

\subsection{Roadmap}
The remainder of this paper is organized as follows. Section II provides an overview of binary analysis techniques and design considerations for systems employing these techniques.  In Section III, commonly exploited vulnerabilities are briefly discussed. Sections IV and V, are detailed discussions of the architecture, techniques and technologies used to implement Mayhem and Mechanical Phish respectively. Section VI compares and contrasts Mayhem's and Mechanical Phish's approach to mitigating path explosion, a common problem that is encountered when using dynamic symbolic execution for path exploration. The last sections contains proposed future research and conclusions.
 
\section{Background}
Despite our best efforts software defects will always exist and given the growing dependence on technology to manage our daily lives, ensuring the safety, security and reliability of software and hardware has become the primary focus of a number of security researchers. Specifically, an emphasis as been placed on binary software analysis, for the simple fact that in many instances only the binaries are available for analysis. This is particularly true when examining embedded firmware, custom operating systems and maleware.

Binary analysis can be difficult because we are missing abstractions provided by programming languages such as data types and data structures. These abstractions make it easier to reason about how data and inputs drive the paths of execution. Despite these challenges there are inherent advantages to performing binary analysis. Binaries contain platform specific details which are only available at execution time.  Information such as ``memory layout, register usage and execution order'' \cite{Balakrishnan2008} is important for detecting many common types of vulnerabilities such as memory corruption and buffer overflows. For these reasons and more, binary analysis a specific type of program analysis is the focus of security researchers in recent years and the volume of software to be examined has lead to a strong interest in building automated binary analysis systems that can examine binary software at scale.

Static, dynamic and concolic analysis (also known as dynamic symbolic analysis) are three common approaches to binary analysis. Each approach has its strengths and limitations and each comes with their own set of design considerations that must be addressed in order to meet the challenge of analyzing real-world software. The following sections examine each of these approaches; their limitations, strengths and the design considerations that must be addressed in order to implement systems that perform automated vulnerability detection and exploit generation effectively.

\subsection{Design Considerations}
One design consideration that must be addressed when implementing automated vulnerability detection and exploit generation systems, is ensuring the ability of the system to replay or reproduce the program state (i.e. user input or data) that triggered a vulnerability. The other consideration is the system must understand semantically what part of a given input caused the observed behavior. These design considerations directly impact the scalability and validity of the results these systems yield (i.e. vulnerabilities discovered).  For example, analysis techniques such as symbolic execution aims to have high reproducibility and high semantic understanding but will suffer from issues with scalability while approaches that favor "re-playability" usually suffer from low code coverage \cite{Shoshitaishvili}.

\subsection{Static Binary Analysis}
\begin{figure}
 \centering
  \includegraphics[scale=0.12]{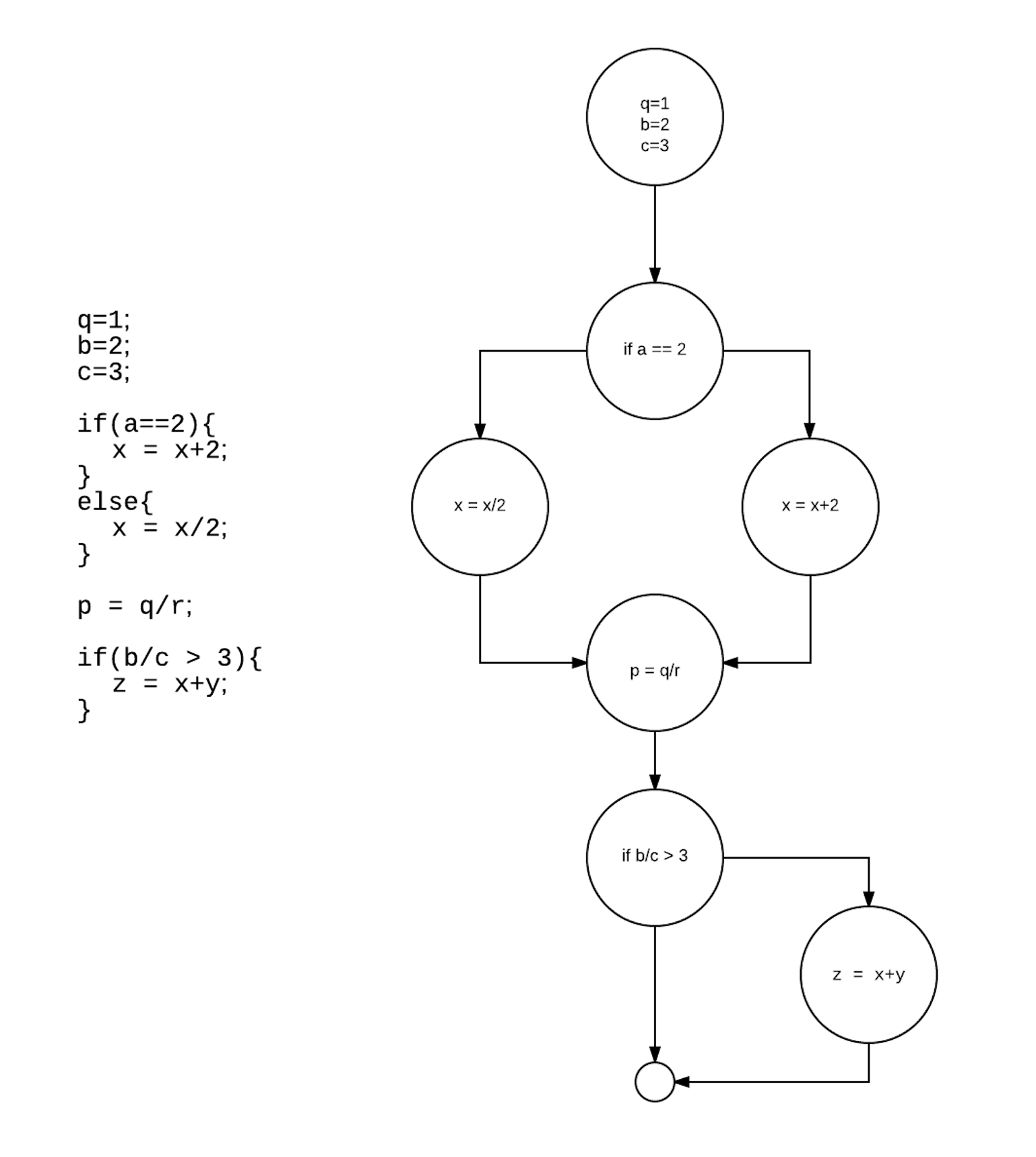}
  \caption{Example of simple control flow graph (adapted) \cite{Copeland2004}}
  \label{fig:simpleCFG}
\end{figure}

Static binary analysis is the analysis of a binary without running it. The process of static binary analysis typically starts with loading and processing the binary to be analyzed. The processing step includes parsing the binary, generating an intermediate language representation of the binary's assembly instructions and building a control flow graph (CFG). Control flow graphs represent paths that can be taken when a program executes. For example, Figure~\ref{fig:simpleCFG} illustrates a simple CFG where program flow is controlled by conditional statements. The nodes of these graphs represent basic blocks of machine instructions and the edges represent possible points of control flow changes between these nodes. Control flow graphs are a key component for automated vulnerability detection systems that employ static binary analysis as it gives the system a means of exploring all execution paths in an application.

\subsubsection{Limitations Of Static Binary Analysis}
Though this technique offers a system the ability to examine all possible program paths, it comes at the cost of scalability and performance. Static binary analysis can be slow, and it has limitations when dealing with indirect jump statements. Indirect jump statements are harder than direct jump statements to resolve when building a CFG because the application is passing control to a target whose value for example, could be arbitrarily calculated or dependent on the context of application.  To deal with these limitations static binary analysis tools make approximations about the control flow of an application and hence run the risk of not resolving indirect jump statements at all. Under approximations can lead to false positives for systems that detect vulnerabilities or worse it could miss detecting vulnerabilities due to incomplete control flow graphs. 

One static analysis technique that mitigates some of these limitations is value-set analysis (VSA). The key to this algorithm is its over approximation of values in memory, a property that makes it useful in making assumptions about targets of indirect jump statements or read and writes in memory. These properties enable VSA to be used to augment CFGs with information about indirect jump statements \cite{Balakrishnan2008}.

\subsection{Dynamic Binary Analysis}
Unlike static binary analysis, dynamic binary analysis techniques examines a program's behavior while it is running in a given environment. Dynamic binary analysis allows you to explore individual paths which makes it very precise but at the expense of less code coverage. Code coverage is an important characteristic of vulnerability detection systems, as the more code you can examine the more likely you are to find existing vulnerabilities.

\subsubsection{Concrete and Symbolic Execution}
There are two flavors of dynamic binary analysis, concrete and symbolic execution. Concrete execution refers to the representation and execution of ``concrete'' or real values against a program, where as symbolic execution refers to the representation and execution of symbolic representations of a given value (i.e. a range of values). In dynamic analysis systems, binaries and source code are augmented with instrumentation \cite{Nethercote2004}, this instrumentation provides metadata to enable the system to make better choices about things like choosing paths in an application to explore.

\subsubsection{Fuzzing}
The main objective of a system that detects vulnerabilities is to find inputs that make it perform an unsafe operation (i.e. crash an application). Fuzzing is an example of concrete execution and it is an important technique used in systems where augmented input is used to attempt to crash an application for example. Though fuzzing is an important technique in vulnerability detection it suffers from limitations. Fuzzing tools usually require manually created test cases to seed the fuzzer. It then mutates its future inputs based on these test cases. Standard fuzzing techniques usually fail to randomly generated values for branches of logic that requires very specific user input or context dependent data. 

\subsection{Dynamic Symbolic Execution}
A more powerful dynamic analysis technique that is implemented in many automated vulnerability detection systems, is dynamic symbolic execution. In classical symbolic execution, variables and application input (i.e. files, command line options etc) are modeled using symbolic values instead of using concrete values. During execution, both memory and register state are tracked and are also modeled symbolically. Symbolic execution is typically used to dynamically generate test cases which are used to drive path exploration, unlike traditional fuzzing techniques where test cases must be manually generated to seed the system. Systems like Mayhem \cite{Cha2012} and S2E \cite{Chipounov2011} were some of the first to apply this technique to binary code. 

In dynamic symbolic execution input and variables are represented as symbolic values instead of concrete values. These values are used to generated path constraints. Path constraints are logical formulas that represent ``program state and transformations between program state'' \cite{Bjorner2011}. Typically these formulas represent previously unexplored paths of execution in a program and are used as input to a satisfiability modulo theory solver (SMT solver). The SMT solver uses these formulas to derive new application inputs (test cases) that drive the exploration of new paths in the application \cite{Bjorner2011}, \cite{Li2014}. Because most programming constructs can be modeled by theories supported by SMT solvers, they are often used in tools that verify and test programs.

\subsubsection{Limitations Of Dynamic Symbolic Execution}
Dynamic symbolic execution is so powerful because it can trigger specific application states using learned path constraints, making it an ideal technique for discovering vulnerabilities in binary code \cite{Shoshitaishvili}. This characteristic makes it a commonly used technique in well known binary analysis tools such as CUTE \cite{Sen2005}, Klee \cite{Cadar2008} and FuzzBALL \cite{Martignoni2012}. However, dynamic symbolic execution suffers from a problem known as path explosion, whereby new paths are created at every new branch. This can lead to an exponential number of paths to be explored and which makes dynamic symbolic analysis computationally expensive, hence limiting the scalability of analysis systems that use this technique as its only mechanism of path exploration. 
 
 A modern approach to combat these limitations is to combine both concrete and symbolic execution, a technique known concolic execution \cite{Cadar2013}. Another approach combines the use of dynamic symbolic execution and fuzzing to create a ``guided'' fuzzer \cite{Stephens2016} or assisted fuzzer. This technique uses dynamic symbolic execution to drive path exploration by giving it the task of augmenting input and feeding it back to the fuzzer. The aim of this technique is to minimize the use of an expensive operation such a dynamic symbolic execution, and use cheaper operations such as fuzzing to get better code coverage when exploring applications for vulnerabilities. This technique is used by Driller \cite{Stephens2016}, where by it selectively uses dynamic symbolic execution to perform path exploration in order to detect vulnerabilities. Driller is key component in the Mechanical Phish Cyber Reasoning System \cite{DARPA}, \cite{Stephens2016}.

\section{Commonly Exploited Vulnerabilities}
Programming languages such as C/C++ gives developers low level control of memory allocation, which allows for finer grain control over application performance and efficiency. This level of control can lead to security critical vulnerabilities that can be exploited by attackers. Although there are efforts to make software more secure and robust with the implementation of techniques such as buffer overflow detection and randomization of address space, vulnerabilities such as buffer overflows are still in the top three vulnerabilities reported in 2015 and 2016 \cite{Stephens2016}\cite{CVEType}. Tables II and III shows the number of reported vulnerabilities for the top three types of vulnerabilities of 2015 and 2016.

Some of the most commonly found exploitable vulnerabilities are buffer overflows, format string attacks and general memory corruption vulnerabilities. These are defects that often put an application in an unsafe state, where an attacker can gain access to sensitive data or hijack the control flow of an application, in order to execute code of their choice. It is for these reasons why most automated vulnerability detection systems seek to detect these types of vulnerabilities. 

\begin{table}
\centering
    \begin{tabular}{ l l }
      Type & Count\\
        \hline
        Denial of Service & 1847\\
        Execute Code & 1355\\
        Overflow & 1221\\
             & \\
    \end{tabular}
     \caption{Top 3 Reported Vulnerabilities by Type (2016) \cite{CVEType}}
\end{table}

\begin{table}
\centering
    \begin{tabular}{ l l}
        Type & Count\\
        \hline
        Denial of Service & 1784\\
        Execute Code & 1808\\
        Overflow & 1072\\
             & \\
    \end{tabular}
      \caption{Top 3 Reported Vulnerabilities by Type (2015) \cite{CVEType}}
\end{table}

\subsection{Buffer Overflows}
Buffer overflows occurs when an application writes more data to a fixed size buffer than it is allocated to handle. Typically these kind of vulnerabilities can lead to data corruption, crashing applications, the unintended access of sensitive data stored in memory or allowing an attacker to replace code in the call stack with their own or a library call of their choice.

\subsection{Format String Attacks}
Format string attacks are used by attackers to execute code or read data from the stack. This exploit occurs when a formatted string given as an input is executed as a command. These kind of attacks often use the ANSI C printf, fprintf and other string format functions as attack vectors.

\subsection{General Memory Corruption}
Buffer overflows are an example of a type of memory corruption vulnerability. Generally memory corruption occurs when data in a previously allocated memory location is modified accidentally or intentionally. The use of this corrupted data can lead to application crashes. Other examples of memory corruption are array index out of bounds errors, using an address before memory is allocated or attempting to use a pointer that has been freed already.

\section{Mayhem}
Mayhem is an automated system for discovering exploitable vulnerabilities in binary code. It also ensures that each vulnerability is exploitable and verifiable by generating a ``shell spawning exploit'' for each vulnerability it finds. By design Mayhem seeks to addresses the challenges of real-world binary analysis by ensuring that it can not only find exploitable bugs but do so efficiently. It does so by employing a technique called hybrid symbolic execution. Mayhem introduces the use of hybrid symbolic execution. Hybrid symbolic execution combines the use of both offline and online symbolic execution \cite{Cha2012}.

Hybrid symbolic execution leverages the strengths of online and offline symbolic execution while minimizing the effects of their limitations.  While offline symbolic execution, also know as concolic execution allows a system to examine one execution path at time while enabling it to select new paths to explore via an iterative process, it has one major limitation. The major limitation of offline symbolic execution is that in order to find new paths, the executor must run a single path of execution twice, once concretely and once symbolically. This re-execution of previously explored paths makes this technique inefficient as it adds additional execution overhead to a system. On the other hand online symbolic execution seeks to execute all paths in a single run and it does so by forking execution at each branch.  Although this approach ensures that the system would never explore a path more than once, the constant forking could lead to memory pressure as all application state is stored in memory. 

The following sections discusses Mayhem's design, some key implementation notes, contributions made by the researchers as well as related work. Note, all information was taken from the literature.

\subsection{System Overview}
The more of an application a vulnerability discovery tool can explore the more likely it is to find exploitable bugs. This presents a major challenge for preforming binary analysis on real-world applications, this can be especially true for common off-the-shelf applications as they can be complex applications with a very large state space to explore. This challenge is one the key motivations behind Mayhem's design.

Mayhem's designers see exploring binary software as a potentially long running process, this is a especially true for running analysis on complex binaries. This means that the system must be able to run for long periods of time while taking care not to exhaust system resources in particular memory. System efficiency is also a motivation behind Mayhem's design. It addresses this by ensuring that no work is ever repeated and that no work is thrown away, all results from a previous analysis should be reusable on other runs \cite{Cha2012}. Lastly, the key principal behind Mayhem's ability to detect vulnerabilities and generate corresponding exploits is that the system must be able to identify where in symbolic memory a load or store address depends on user input \cite{Cha2012}. 

\subsection{Architecture}
Mayhem's architecture is comprised of two major components, a ``Concrete Executor Client (CEC)`` and a ``Symbolic Executor Server (SES)``\cite{Cha2012}. The SES is the brains of the operation as it determines the next path the CEC should explore and the CEC is the worker, it performs path exploration. The CEC runs natively on the target system and the SES runs independently on the platform's system.

\subsubsection{Concrete Executor Client (CEC)}
The CEC takes symbolic input sources, the binary to be analysed and checkout point data (optional) as input. Because symbolic execution is slower than native (concrete) execution Mayhem seeks to perform as much native execution as possible. The CEC performs the task of loading and natively executing the binary to be analysed. As the binary is executed the CEC adds instrumentation to the code, this instrumentation adds information about execution state such as memory and register values. The CEC also contains a taint tracker which performs taint based analysis \cite{Schwartz2010}. If the taint tracker in the CEC detects a condition or jump statement it halts execution and passes this information to the SES \cite{Cha2012}. Note, the CEC will run until all execution paths have been explored or a threshold is reached.

\subsubsection{Symbolic Executor Server (SES)}
The SES takes the concrete, ``tainted instructions'' from the CEC. These instructions can be a tainted branch or tainted jump instructions. These instructions are converted from x86 assembly to an intermediate language called BAP IL, by BAP. BAP is a binary analysis tool that converts x86 assembly into an intermediate language \cite{Brumley2011}.  The SES takes these interpreted instructions and executes them symbolically. These instructions are used to build two types of formulas, path formulas which represents the constraints on ``each line of code'' and exploitable formulas, which are used to determine if an attacker can execute a payload or gain control of a pointer \cite{Cha2012}. These formals are executed by an SMT solver \cite{Li2014}, which determines if the formula is satisfiable.

To manage system resources Mayhem makes use of configurable resource caps and system generated checkpoints. If a resource cap is not reached and the SES receives a tainted branch instruction, the SES queries the SMT solver to determine if should fork execution. If it forks, a path selector prioritizes the new forks and the SES alerts the CEC about the state change. However, if a system resource cap has been reached then a checkpoint manager generates a new checkpoint for the active executor instead of forking new executors. Note, checkpionts store symbolic execution state of the executor that has been suspended as well as corresponding path constraints. Checkpoint restoration basically uses the stored symbolic execution state to restore the concrete execution state up to the point where the corresponding executor was suspended. Checkpoint restoration essentially puts system back into ``online'' mode. Throughout the execution process the SES switches between existing forked executors and checkpoints \cite{Cha2012}.

\subsection{Minimizing Search Space and Path Selection}
One major challenge of cyber reasoning systems is the vast size of the execution path search space. To address this, Mayhem also uses as technique called preconditioned symbolic execution \cite{ThanassisAvgerinosSangKilCha2012}, which allows the user to provide ``partial specification of the input'' (i.e. input length, prefix etc) \cite{Cha2012} in order to minimize the search space. If no specifications are supplied, all paths will be explored.

Mayhem uses heuristics to determine which path to explore next. It favors paths that are more likely to have an exploitable bug. Paths where symbolic memory accesses occur or symbolic instruction pointers are identified, have higher priority than paths that are simply exploring new paths \cite{Cha2012}. These priority ranking rules directly corresponds to the the types of vulnerabilities that Mayhem (as of the time the initial literature was published) can identify.

\subsection{Handling Symbolic Memory}
Being able to identify where in symbolic memory a load or store address is that depends on user input is a necessity when generating exploits. In order to identify these addresses in symbolic memory, a binary analysis system must be able to model and reason about symbolic memory \cite{Cha2012}.  Modeling symbolic memory is difficult because the index used in the memory look up is an expression instead of a number, this makes dealing with symbolic indices difficult because the index could point to any spot in memory. To tackle this problem Mayhem implements ``index-based memory modeling''. In this approach memory is modeled as a map, and 32 bit indices are mapped to expressions and only symbolic reads are modeled symbolically.

Mayhem uses immutable ``memory objects,'' to model symbolic reads. These objects are created every time a symbolic read is executed, and contain all possible values that the given symbolic index can access. In order to create these objects Mayhem must find all possible values for a symbolic index. In order to make this process more scalable it finds a range of possible index values instead of trying to find an exact index value \cite{Cha2012}. It uses an SMT solver to resolve this range of values. Querying the SMT solver for a range of symbolic index values is an expensive operation, so as an optimization step Mayhem first uses value-set analysis \cite{Balakrishnan2008} to come up with an approximate interval of possible index values, which is then given to a SMT solve to refine or ``tighten`` the lower and upper bounds \cite{Cha2012}.

\subsection{Generating Exploits}
Mayhem (as of the time the initial literature was published) can identify and generate exploits for any ``instruction-pointer overwrite'' and format string attacks. It generates an exploitable formula whenever its exploitable properties are violated. These properties are a symbolic tainted instruction pointer, which corresponds to a buffer overflow and a symbolic format string which corresponds with a format string attack  \cite{Cha2012}.

\subsection{Related Work}
Mayhem's approach to exploitable bug discover and exploit generation is largely based on prior work by the researchers on AEG \cite{ThanassisAvgerinosSangKilCha2012},  \cite{Cha2012}. Unlike Mayhem, AEG used source code analysis to find exploitable bugs while using binary runtime information to generate corresponding exploits. AEG was the first system to provide an end to end solution that not only detects exploitable bugs but generates a verifiable exploit to confirm that it is a security risk. Its approach to automatically generating exploits addressed the issue that source code alone can not tell you if a bug is exploitable. Source code though it provides useful abstractions does not provide the same low level details that are a necessity in determining if a bug can be exploited. AEG also introduced the use of preconditioned symbolic execution to minimize search space as well path prioritize heuristics \cite{ThanassisAvgerinosSangKilCha2012}.

\section{Mechanical Phish}
Mechanical Phish is an open source Cyber Reasoning System written for the DARPA Cyber Grand Challenge \cite{DARPA}. It leverages open source tools and is comprised of several components that directly contribute to its approach to vulnerability discovery and exploit generation.  Mechanical Phish's goal is to discovery vulnerabilities deeper in binary code efficiently. It does so by employing the use of a ``guided'' fuzzer which combines the efficiency of fuzzing and concolic execution with the power of dynamic symbolic execution. 
The following sections discuss Driller's and angr's design, some key implementation notes, contributions made by the researchers as well as related work. Note, all information was taken from the literature.

\subsection{System Overview}
Two important components that implement Mechanical Phish's vulnerability discovery functionality are angr \cite{Shoshitaishvili} and Driller \cite{Stephens2016}. Driller is a ``guided whitebox fuzzer'' tool that leverages the speed of fuzzing and the input reasoning capabilities of concolic execution in order to effectively and efficiently discover deeper bugs. angr is an open source binary analysis framework that Driller uses to implement its concolic execution engine.

\subsection{Driller}
Driller's primary objective is to find bugs in the deeper logic of any application. This objective is the motivation for its approach of leveraging the strengths of fuzzing and concolic execution while mitigating their weaknesses. Systems that implement fuzzing or concolic execution alone, are often limited in the depth and the amount of code they cover because of the inherent limitations of fuzzing and concolic execution \cite{Stephens2016}. Traditional fuzzing techniques are fast but fail to find bugs where specific input is required, while concolic execution is a great tool to generate this kind of input it often suffers from the path explosion problem. By combining these techniques Driller can improve the scalability of concolic execution while also improving the effectiveness of fuzzing. Unlike some systems that only support discovery of specific types of vulnerabilities, Driller can detect any vulnerability that can lead to an application crash.

The core motivation behind Driller's design is that it views the types of bugs that fuzzing and conconlic execution can find in terms of how an application processes input. It splits the input processed by an application into two categories, general and specific. General input can represent a wide range of valid values while specific input only can only have a small number of valid values. This intuitively splits the application into ``compartments,`` where the specific input checks separates one compartment from another. Because fuzzing is an effective technique for generating values for general inputs it can be used to explore application paths within a compartment, while concolic execution would be best used to resolve inputs to drive code execution between application compartments \cite{Stephens2016}.

Driller is comprised of two major components, the fuzzer and concolic execution engine. The bulk of the path exploration work is offloaded onto the fuzzer as in most cases it can explore a large number of execution paths on its own. This leaves the concolic execution engine to solve for the more complex inputs required by specific checks in an application.

\subsubsection{The Fuzzer}
The fuzzer component leverages a very popular fuzzer called American Fuzzy Lop (AFL) \cite{AFL}. AFL is a state of the art fuzzer that generates input through the use of a genetic algorithm. It uses instrumentation to make more informed choices. Though instrumentation can be introduced at compile time, Driller uses a ``QEMU-backend`` \cite{Bellard2005} to avoid the need for having source code.  The bulk of the path exploration work is offloaded onto the fuzzer. In most cases it can explore large number of execution paths, and is much faster than concolic exeuction.  

\subsubsection{Concolic Execution Engine}
The concolic execution engine uses angr \cite{Shoshitaishvili}, an open source binary analysis framework. This engine, translates binary code into Valgrind's VEX \cite{Nethercote2007} intermediate representation (IR). This IR is used to evaluate the effects of application input on symbolic state. All values in the symbolic state except constants are modeled as symbolic variables and as the program is executed ``symbolic constraints'' are added to the symbolic variables. These constraints defines the limit of possible values for a symbolic variable. Throughout execution both concrete and symbolic values are tracked, these values can be used by the constraint solver (SMT solver) to find values that satisfy the constraints on all symbolic variables in the state \cite{Stephens2016}. Like Mayhem, Driller uses the index-based memory model to model symbolic memory where writes addresses are stored concretely and read addresses are modeled symbolically \cite{Cha2012}.

When Driller starts it invokes the fuzzer component. The fuzzer will explore the application until it can no longer generate inputs that drive execution down new paths. When the fuzzer reaches this state Driller says the fuzzer is ``stuck.'' More concretely the fuzzer is deemed stuck if after having gone through a certain number of input mutations it fails to progress to new paths. When the fuzzer is stuck, Driller takes only the inputs the fuzzer marks as ``interesting'' and invokes the concolic execution engine on them. Inputs are considering interesting if the input triggers a state transition \cite{Stephens2016}.

\subsection{angr}
angr \cite{Shoshitaishvili} is an open source, platform agnostic binary analysis framework, that implements a number of state of the art offensive binary analysis techniques. This framework was implemented to provide researchers a unified platform by which they can evaluate and compare the effectiveness of these techniques as well as components to implement and evaluate new techniques.

\subsubsection{Motivation}
Many binary analysis techniques are developed as research prototypes and are typically not available to the public.  This often means that future researchers have to start from scratch in order to implement and evaluate these techniques themselves \cite{Shoshitaishvili}. angr was created to mitigate this issue, by creating an open source, binary analysis framework that implements the state of the art offensive binary analysis techniques. The framework also offers a modular design that allows researchers to easily combine different approaches in a effort to leverage their strengths while minimizing their weaknesses.

angr is implemented as collection of Python libraries. The libraries provide functionality for performing various binary analysis techniques:

\begin{itemize}
\item Loading a binary
\item Disassembly and intermediate-representation lifting
\item Program instrumentation
\item Symbolic execution
\item Control-flow analysis
\item Data-dependency analysis
\item Value-set analysis (VSA)
\end{itemize}

\subsection{angr Submodules}
angr's primary design goals are to offer cross-architecture support, cross-platform support, support for different analysis techniques, and usability. With these goals in mind the researchers that created angr wanted to create a system that would allow users to recreate any common binary analysis technique in about a week. In order to accomplish these goals, the analysis engine was carefully designed to be a modular set of software components with strict separations between them. This design allows for the mixing and converting between types of analysis on-the-fly \cite{Shoshitaishvili}.

The sections below provide a brief summary of some key submodules implemented in angr.

\subsubsection{CLE}
CLE is angr's binary loading module. It can support loading binaries from POSIX-compliant systems such as Linx, FreeBSD, Windows as well as DECREE OS which was created by DARPA for the Cyber Grand Challenge.

\subsubsection{Intermediate Representation}
In order to support analyzing binaries in a architecture agnostic way it is necessary to convert binary code into an intermediate representation (IR). The module that supports IR in angr leverages libVEX and uses a python library called PyVex to expose libVEX's VEX IR in python. PyVex, was originally written for Firmalice \cite{Shoshitaishvili2015}. VEX allows angr to support analysis of both ``32-bit and 64-bit versions of ARM, MIPS, PPC, and x86 (with the 64-bit version of the latter being amd64) processors'' \cite{Shoshitaishvili}.

\subsubsection{SimVex}
Functionality for representing and modifying program state is implemented in the SimVex module. State (SimState) in angr is represented by collection of ``state`` plugins. These state plugins provide the building blocks for implementing different types of binary analysis. These plugins expose functionality for; tracking values of registers, implementing symbolic memory modeling, implementing abstract memory modeling, logging, debugging, providing an interface for interacting with SMT solvers, and exposing architecture specific information that is useful for analysis.  

\subsubsection{Claripy}
Claripy is the module responsible for provide abstractions that represent values stored in SimState. Claripy internally represents these values as expressions that can be translated to the data domains of various supported Claripy back-ends. Claripy supports back-ends for concrete domains, symbolic domains and value-set abstraction domains for value-set analysis.

\subsubsection{Program Analysis}
angr implements complete analysis techniques such as dynamic symbolic execution and control-flow graph recovery. It exposes an entry point that allows users to easily access all things related to the analysis, such as the binary being analyzed and exposes functionality of various submodules.

\subsubsection{Other Key System Components}
This section is a brief discussion of other key software components of angr. Note, all documentation below was obtained from the Mechanical Phish github repository \cite{MechaphishRepository}.

\begin{itemize}
\item \textbf{Rex:} Rex is an automated exploitation engine that was originally implemented for the Cyber Grand Challenge.  As of the time of writing this paper the engine can perform crash triaging, crash exploration, and exploitation for certain kinds of crashes. Rex is freely available on github \cite{RexEngine}
\item \textbf{Meister:} Meister is the task scheduler for Mechanical Phish.
\item \textbf{Scriba:} Scriba decides what exploits and replaceable binaires (CGC patched binaries) to submit.
\item \textbf{The Ambassador:} The Ambassador talks to the CGC API to retrieve challenge binaries, submit proof of vulnerabilities, etc.
\end{itemize}

\subsection{Minimizing Search Space and Path Selection}
To avoid the problem of path explosion in the concolic execution engine, Driller implements ``pre-constrained tracing.'' Pre-constrained tracing ensures that the only path that is being analyzed is the path that represents the application's processing of a given input \cite{Stephens2016}.

\section{Compare \& Contrast}
This survey explored the current state of the art offensive capabilities implemented in Cyber Reasoning Systems. It used two of the winning systems, Mayhem and Mechanical Phish of DARPA's Cyber Grand Challenge as a vehicle to explore these techniques in action. The original motivation for this survey was to investigate the similarities and differences of these two systems in order to identify what sets them apart and which approaches worked best, for solving the various problems that must be addressed in order to build an automated system that can successfully detect exploitable bugs and generating exploits for these bugs. This section provides a brief discussion of some key differences between Mayhem and Mechanical Phish.

\subsection{Path Explosion}
Mayhem and Mechanical Phish both leverage dynamic symbolic execution in order to drive path exploration but their approaches to mitigating the problem of path explosion are different. Dynamic symbolic execution is a popular technique for discovering vulnerabilities in binary code, and it works well finding both both complex and simple inputs to drive path exploration.  However, dynamic symbolic execution suffers from a well known problem of path explosion, where by new paths are created at every new conditional branch. This can lead to an exponential number of paths to be explored and hence makes dynamic symbolic analysis computationally expensive and can limit the scalability of analysis systems that use this technique as its only mechanism of path exploration. Hence, any system looking to employ this technique must address the problem of path explosion.

\subsubsection{Hybrid Symbolic Execution}
Mayhem uses dynamic symbolic execution as its primary mechanism of implementing path exploration. In order to combat the path explosion problem, it implements hybrid symbolic execution. Hybrid symbolic execution allows the system to switch between online and offline symbolic execution. Its ability to context switch between offline and online executors allows it to use to the most appropriate mechanism whenever a configurable resource cap is reached. This allows Mayhem to technically have its cake and eat it too, as it can leverage a powerful method of path exploration without succumbing to its limitations.

\subsubsection{Augmenting A Fuzzer With Symbolic Execution}
Mechanical Phish uses Driller to help with path exploration. Driller's approach to avoiding the pitfall of symbolic execution by using its fuzzer to perform the bulk of path exploration and only leveraging symbolic execution when the fuzzer ``gets stuck'' or in other words fails to generate an input that can drive path exploration forward.  The key to this approach is its use of pre-constrained tracing which ensures that the only path that is being analyzed is the path that represents the application's processing a given input \cite{Stephens2016}.

\section{Proposed Future Research}
This section will briefly discuss some proposed areas of future research.  

\subsection{Binary Pre-processing To Minimize Search Space For Large \& Complex Applications}
In many ways binary analysis can be viewed as an uninformed search problem, that when coupled with tools like instrumentation it evolves into an informed search problem. The search space for large, complex applications can be vast and systems that are seeking to effectively perform analysis on such applications at scale, must find even more effective ways (than the current state of the art) to minimize the search space. Minimizing the search space leads to be better code coverage which enables analysis tools to find defects deeper in code. 

With the above issue in mind, I propose exploring creating a system that can perform binary pre-processing, with the purpose of identifying application ``hot spots.'' Hot spots are areas in an application where exploitable bugs are likely to exist. These hot spots would split an application into regions. Information regarding these hot spot regions, path constraints and other metadata, would act as a map or guide to that area in the code. This information would be given along with the corresponding binary to a vulnerability detection system, and this system would use this metadata and path constraints to make its way directly to the hot spot region. Once the system reaches this region, it would perform binary analysis as normal. This pre-processing step could make vulnerability detection an even more informed search problem, and by splitting software into regions it would give less code to reason about at one time during the vulnerability detection phase.

\subsection{Generating Exploits From Common Vulnerabilities and Exposures Reports (CVEs)}
A human security analyst or attacker has the ability to read a Common Vulnerabilities and Exposures report (CVE) and build exploits for the reported vulnerability. They are leveraging not only the knowledge supplied in the CVE, they are also leveraging their historic knowledge and past experience to generate an exploit for the reported vulnerability. I propose exploring the feasibility of creating a knowledge based system that leverages the analysis capabilities of the state of the art vulnerability discovery and exploit generation tools to learn the common characteristics of exploits and vulnerabilities. This system would take these learned insights and attempt to generate a generic exploit for a given common vulnerability report. This would create a system that doesn't require the source code or binary code to generate test cases (exploits) for a given vulnerability it would only need the binary or source code to verify the test cases it generates.

\subsection{Deep Reinforcement Learning For Vulnerability Discovery}
Cyber Reasoning Systems are expert systems that encapsulate the actions and knowledge of a human analyst in an automated system that can detect exploitable bugs, generate verifiable exploits, and patch software.  Binary code shares similar characteristics as a board or world in a video game, as they both are subject to state changes based on user defined input and interactions. Research in the field of deep reinforcement learning has proven that an intelligent software agent is capable of learning and excelling at complex tasks \cite{Mnih2015}. I propose exploring the feasibility of combining the use of a binary analysis system with deep reinforcement learning to create an AI agent that can learn to discover vulnerabilities in binary code.

\section{Conclusion}
Cyber Reasoning Systems are expert systems that encapsulate the actions and knowledge of a human analyst in an automated system that can detect exploitable bugs, generate verifiable exploits, and patch software. These systems are complex and require expert knowledge of the problem to build them.  Though there are still many open problems that need to be addressed in order for Cyber Reasoning Systems to be able to reason about real-world complex applications, the current state of the art systems prove that it is very possible to build automated systems that can perform automated vulnerability detection, exploit generation and software patching in binary software without human intervention.

\medskip
 
\bibliographystyle{IEEEtran}
\bibliography{tbrooks_survey_of_crs}

\end{document}